\documentclass[12pt,Letter]{article}
\usepackage{amssymb}
\usepackage{amsfonts}
\usepackage{amsmath}
\usepackage{graphicx}
\usepackage{appendix}
\usepackage[left=2cm,top=2cm,right=2cm,bottom=2cm]{geometry}
\usepackage{epstopdf}
\usepackage{float}

\input{tcilatex}

\begin{document}

\title{{\Huge On the Klein's paradox in the presence of a scalar potential}}
\author{B Ainouz and S. Haouat \\
\textit{LPTh, Department of Physics, }\\
\textit{Mohamed Seddik Ben Yahia University (Jijel),}\\
\textit{BP 98, Ouled Aissa, Jijel 18000, Algeria.}}
\maketitle

\begin{abstract}
In this paper, we have studied the Klein's paradox in the presence of both
scalar and vector potential barriers. From the corresponding Dirac equation
we have calculated the transmission and reflection coefficients. It is shown
that the presence of a scalar barrier the scalar potential wides the gap
between positive and negative energies and so the forbidden region.
Accordingly, the Klein's paradox disappears when the scalar barrier exceeds
a critical value. Considering the problem within the framework of quantum
field theory, we have calculated the related pair creation probability, the
mean number of created particles and the probability of a vacuum to remain a
vacuum. Then it is shown that the scalar potential cut down the Klein range
and minimizes the creation of particles; The particle creation decreases as
the scalar potential increases and ceases definitely when the scalar
potential reaches the critical value.


\end{abstract}

\section{Introduction}

It is widely known that relativistic effects influence the quantum behavior
of particles and can significantly modify the results of nonrelativistic
quantum mechanics \cite{Bjork,Grein2,Bagr}. The most surprising result of
relativistic quantum mechanics is the famous Klein's paradox \cite{Klein}.
Briefly, this paradox is associated with the scattering of relativistic
particles by a potential barrier where the flux of the reflected wave by the
barrier may be greater than that of the incident wave \cite{Calo}. This
result has found a theoretical interpretation after the introduction of the
Dirac's hole theory which predicted the existence of antiparticles
associated with particles. According to this theory, the physical vacuum is
not really "empty" but a "Dirac sea" which contains virtual particles having
a negative energy. Then, the scattering of particles is accompanied by the
transition of the virtual particles of the Dirac Sea into real particles by
the tunnel effect. This transition produces particle-antiparticle pairs in
the vicinity of the barrier. Of course, this transition can take place only
for a potential barrier greater than twice of the particle mass.

In literature, we can find a big number of papers addressed to the Klein's
paradox and associated particle creation where different shapes of the
barrier are considered. Namely, the step potential \cite{Klein}, the square
barrier \cite{Xue}, Sauter type potential \cite{Sauter,Nikish1} and the
inhomogeneous $x$-electric potential steps \cite{Git2}. Some authors
considered also different wave equations such as Klein Gordon equation \cite%
{Nikishov}, Feshback-Villars equation \cite{Bounames,hao,Bounames2}, Dirac
equation \cite{Nikish1} and the DKP equation \cite{Chetouani,Merad2,Cardoso}%
. However, except of few works considering nonminimal coupling \cite%
{ASdecastro,hao1,Haouat}, the case studied in detail is that of charged
particles coupled minimally to a vector potential.

In the present paper, we discuss the Klein's paradox in the presence of a
scalar potential by considering the Dirac equation with vector and scalar
potential barriers. Since the scalar potential couples to the particle mass,
it modifies the gap between positive and negative energies and,
consequently, may have an important effect on the Klein paradox. This could
be of interest to modern physics for the reason that the experimental
observation of the related particle creation has been regarded as a
challenge for quantum electrodynamics. It should be noted that the
scattering of relativistic particles and associated Klein paradox are
discussed, in a certain way, in \cite{Castilho,Castilho1,Castilho2}. In the
present work, we concentrate our attention to the effect of the scalar
potential on the particle creation in the vicinity of the barrier by
considering two approaches; a direct calculation based on relativistic
quantum mechanics and rigorous treatment in the framework of quantum field
theory.

The paper is organized as follows; At the beginning, we solve the Dirac
equation with a general mixing of vector and scalar potentials and we
discuss the appearance of the Klein's paradox. Next, we propose to treat the
problem within the framework of quantum field theory where we give two sets
of exact solutions that can be interpreted as "in" and "out" states. Then,
by the use of the relation between these two sets we determine the
probability of pair creation, the mean number of created particles and the
vacuum persistence.

\section{Klein's paradox in the presence of a scalar potential}

Let us consider a Dirac particle with mass $m$ subjected to a generalized
potential containing both the usual vector potential and a Lorentz scalar
potential. The dynamics of this particle is in general governed by the
stationary Dirac equation

\begin{equation}
E\Phi =\left[ \boldsymbol{\alpha }p+\boldsymbol{\beta }\left( m+S(z)\right)
+V(z)\right] \Phi  \label{2}
\end{equation}%
where $\boldsymbol{\alpha }$ and $\boldsymbol{\beta }$ are the Dirac
matrices. In (1+1) dimensional space-time, $\boldsymbol{\alpha }$ and $%
\boldsymbol{\beta }$ are given by the ($2\times 2$) representation%
\begin{equation}
\boldsymbol{\beta }=\sigma _{z}~\text{~},\text{~}\boldsymbol{\alpha }=\sigma
_{x},
\end{equation}%
where~$\sigma _{x}~$and $\sigma _{z}~$are the Pauli matrices
\begin{equation}
\sigma _{x}=\left(
\begin{array}{cc}
0 & 1 \\
1 & 0%
\end{array}%
\right) ,~\ \ \ \ \sigma _{z}=\left(
\begin{array}{cc}
1 & 0 \\
0 & -1%
\end{array}%
\right) .
\end{equation}%
Now, for simplicity reasons, we consider the following scalar and vector
barriers

\begin{equation}
V(z)=\frac{V_{0}}{S_{0}}S(z)=\left\{
\begin{array}{c}
V_{0}\left. {}\right. if\left. {}\right. z>0 \\
0\left. {}\right. if\left. {}\right. z<0%
\end{array}
\right.  \label{V}
\end{equation}
The definition (\ref{V}) divides our space on two regions. The first one is
the left region ($z<0)$, where the particle behaves like free one. The
second region is to right of the barrier. In this region the particle has a
an effective mass due to the scalar potential.

Following the general consideration, there is two solutions to equation (\ref%
{2}) in the left region ($z<0$). It's about the incident and reflected waves

\begin{equation}
\Phi _{\rightleftarrows }(z)=\sqrt{\frac{E+m}{2p}}%
\begin{pmatrix}
1 \\
\frac{\pm p}{E+m}%
\end{pmatrix}%
\exp (\pm ipz)  \label{3}
\end{equation}%
with
\begin{equation}
p=\sqrt{E^{2}-m^{2}}.
\end{equation}%
In the right region $(z>0)$, we have only one solution
\begin{equation}
\Phi (z)=\sqrt{\frac{\left( E-V_{0}\right) +\left( m+S_{0}\right) }{%
2\left\vert q\right\vert }}%
\begin{pmatrix}
1 \\
\frac{q}{\left( E-V_{0}\right) +\left( m+S_{0}\right) }%
\end{pmatrix}%
\exp (iqz)  \label{4}
\end{equation}%
with
\begin{equation}
q=\sqrt{\left( E-V_{0}\right) ^{2}-\left( m+S_{0}\right) ^{2}}.  \label{q}
\end{equation}%
We can then write the\ general solution as
\begin{align}
\Phi (z)& =\frac{1}{\bar{t}}\sqrt{\frac{E+m}{2p}}\left[
\begin{pmatrix}
1 \\
\frac{p}{E+m}%
\end{pmatrix}%
\exp (ipz)+\bar{r}%
\begin{pmatrix}
1 \\
\frac{-p}{E+m}%
\end{pmatrix}%
\exp (-ipz)\right] \\
\Phi (z)& =\sqrt{\frac{E-V_{0}+m+S_{0}}{2\left\vert q\right\vert }}%
\begin{pmatrix}
1 \\
\frac{q}{E-V_{0}+m+S_{0}}%
\end{pmatrix}%
\exp (iqz),
\end{align}%
where $\bar{r}$ and $\bar{t}$ are the amplitudes of the reflected and
transmitted waves, respectively. As in the ordinary quantum mechanics, the
continuity condition of the wave function and its derivative at $z=0$, gives
us the equation defining the amplitudes $\bar{r}$ and $\bar{t}$
\begin{align}
\left( 1+\bar{r}\right) & =\frac{\bar{t}}{\sqrt{\left\vert \gamma
_{s}\right\vert }} \\
\left( 1-\bar{r}\right) & =\bar{t}\sqrt{\left\vert \gamma _{s}\right\vert }
\end{align}%
with
\begin{equation}
\gamma _{s}=\frac{\left\vert q\right\vert }{p}\frac{E+m}{E-V_{0}+m+S_{0}}
\label{gam}
\end{equation}%
We easily get%
\begin{equation}
\bar{t}=\frac{2\sqrt{\left\vert \gamma _{s}\right\vert }}{\left( 1+\gamma
_{s}\right) }~\ \ \ \ \ \ \ \ \ \ \ \bar{r}=\frac{1-\gamma _{s}}{1+\gamma
_{s}}.
\end{equation}%
Here, we distinguish two situations. The first case is when $V_{0}>2m+S_{0}$
and the second one is when $2m+S_{0}>V_{0}>2m$.
\begin{figure}[H]
\centering
\includegraphics[width=110mm]{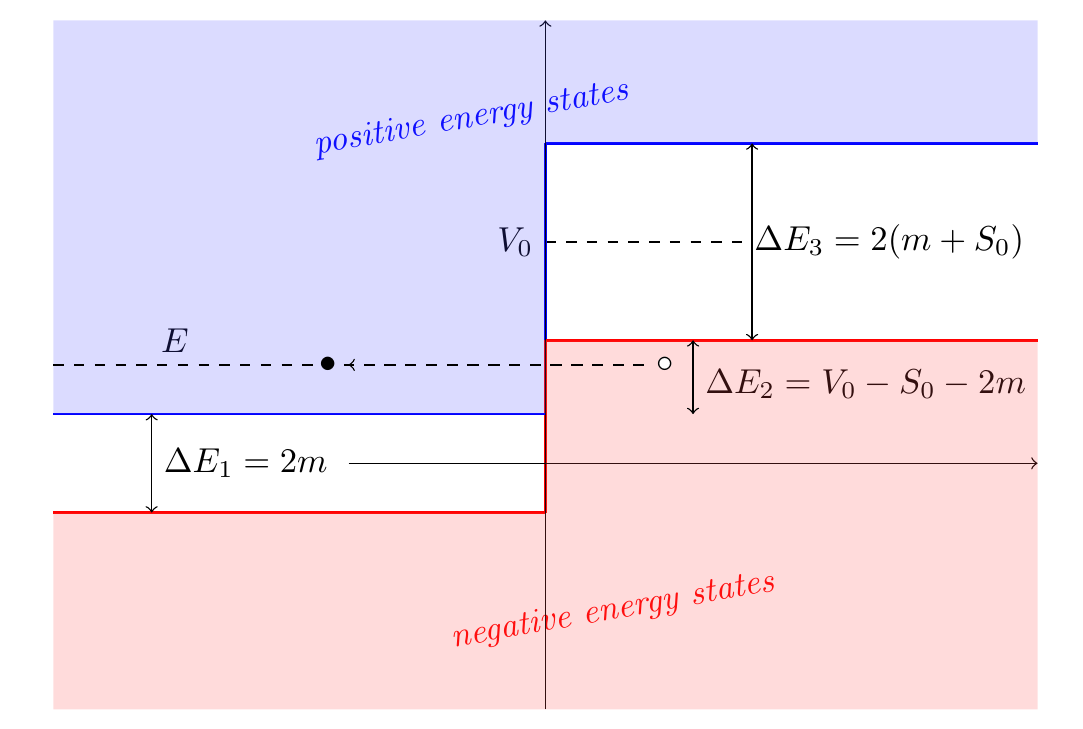}
\caption{The gap between positive and negative energies for $V_{0}-S_{0}>2m$%
. To the left of the barriers, the energy E is located in the domain of
positive energies. To the right of the steps, E is located in the range of
negative energies. This leads to the Klein's paradox. }
\label{fig:g1}
\end{figure}
Let us first consider the case with $V_{0}>2m+S_{0}$. In this case, we have
three ranges for the energy $E$.

\begin{enumerate}
\item The first range is defined by $E>V_{0}+m+S_{0}$. As long as $%
E>V_{0}+m+S_{0}$, the parameter $\gamma _{s}>0$ and we have
\begin{equation}
R=\left\vert \frac{j_{ref}}{j_{inc}}\right\vert =\left\vert \bar{r}%
\right\vert ^{2}=\left( \frac{1-\gamma _{s}}{1+\gamma _{s}}\right) ^{2}
\end{equation}%
and%
\begin{equation}
T=\left\vert \frac{j_{trans}}{j_{inc}}\right\vert =\left\vert \bar{t}%
\right\vert ^{2}=\frac{4\left\vert \gamma _{s}\right\vert }{\left( 1+\gamma
_{s}\right) ^{2}}
\end{equation}%
Here, we can check that $R<1$ and%
\begin{equation}
R+T=1
\end{equation}

\item The second range is when $V_{0}-m-S_{0}<E<m+S_{0}+V_{0}$.\ In such a
case, the wave vector $q$ to the right of the barriers is imaginary and the
solution decays exponentially. In particular, when $E-V_{0}<<m$, then the
solution $\exp (-\left\vert q\right\vert z)$ is localized to within a few
Compton wavelengths. We have complete reflection with exponential
penetration into the classically forbidden region.

\item The third range is the so-called Klein's zone, when $V_{0}-m-S_{0}>E>m$%
. Here, according to (\ref{q}), $q$ becomes real and we obtain an
oscillating transmitted plane wave. This is the first manifestation of the
Klein paradox. This surprising result is due to the fact that the solutions
with $V_{0}-m-S_{0}>E>m$ are positive energy solutions in the first region
and negative energy solution in the second region. Consequently, instead of
complete reflection with exponential penetration into the classically
forbidden region, we have a transition into negative energy states with $%
E<V_{0}-m-S_{0}$.

If the potential is strong enough, $V_{0}>2m+S_{0}$, and $E<V_{0}-S_{0}-m$
the parameter $\gamma _{s}$ becomes negative, $\bar{\gamma}<0$, and the
reflected current would be then greater than the incident current.
Consequently, the flux going out to the left exceeds the incoming flux and
we have%
\begin{equation}
R-T=1.
\end{equation}%
This is an other manifestation of the Klein's paradox \cite{Schwabl}.

Following Feynman's interpretation, the antiparticles are moving backward in
time. Then, the wave function $\psi \approx \exp \left( iqz\right) =\exp %
\left[ -i\left( -q\right) z\right] $\ \ describes an antiparticle that moves
to the left region \cite{Holst}. This represents a total reflection of the $%
"in"$ (incoming) particle of the potential barrier accompanied by
particle-antiparticle creation. It is said that pairs are created in the
vicinity of the barrier.
\end{enumerate}

\begin{figure}[H]
\centering\includegraphics[width=110mm]{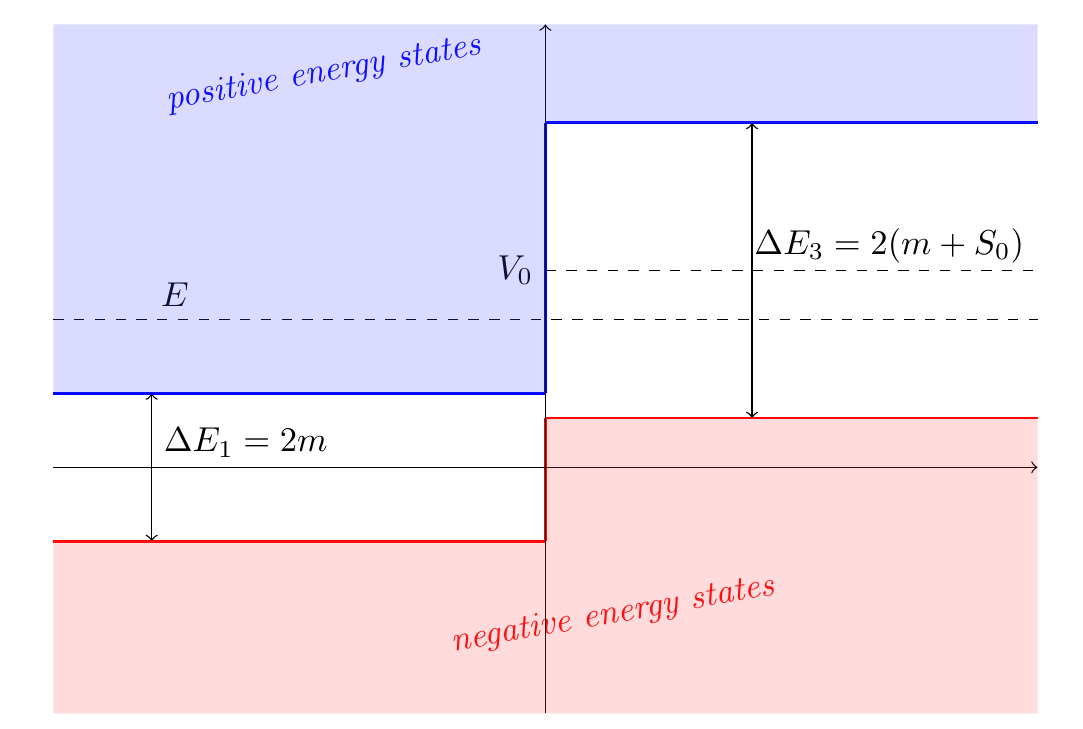}
\caption{The gap between positive and negative energies for $V_{0}-S_{0}<2m$%
. To the left of the barriers, the energy E is located in the domain of
positive energies. To the right of the steps, E is located in the forbidden
region. The solution is then exponentially decaying.}
\end{figure}
Let us note that in the case when $2m+S_{0}>V_{0}>2m$, there is no overlap
between positive energy states in the first region and negative energy
states in the second region. In such a case, we have either a complete
reflection with exponential penetration in the forbidden region or a
transition from a positive energy state in the first region to a positive
energy state in the second region. Therefore, even if $V_{0}>2m$, Klein's
paradox can not take place as long as $V_{0}<2m+S_{0}$.

\begin{figure}[H]
\centering
\includegraphics[width=110mm]{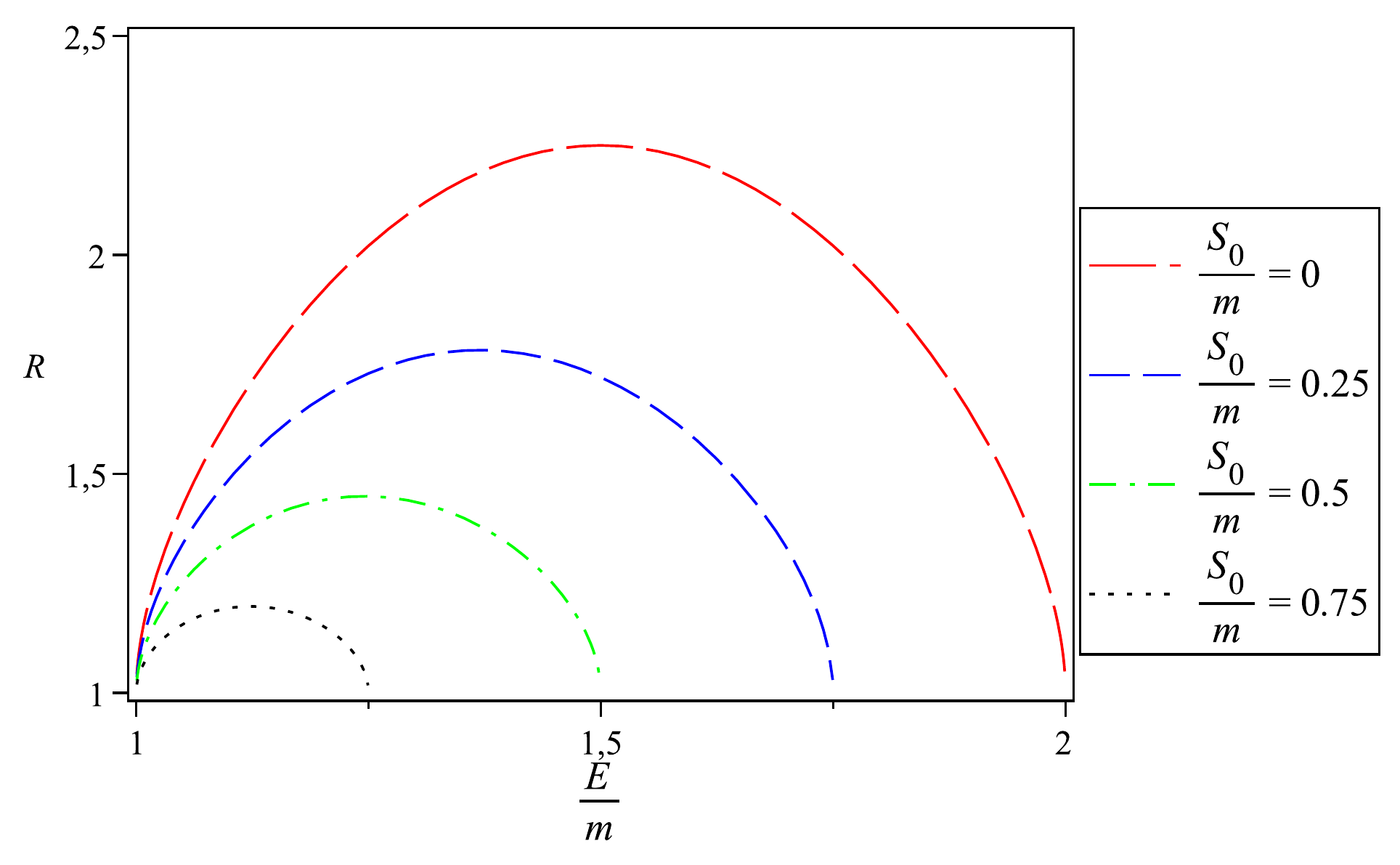}
\caption{The variation of $R$ versus $\frac{E}{m}$. $V_{0}$ is taken $%
V_{0}=3m$.}
\label{fig:g3}
\end{figure}
\begin{figure}[H]
\centering
\includegraphics[width=110mm]{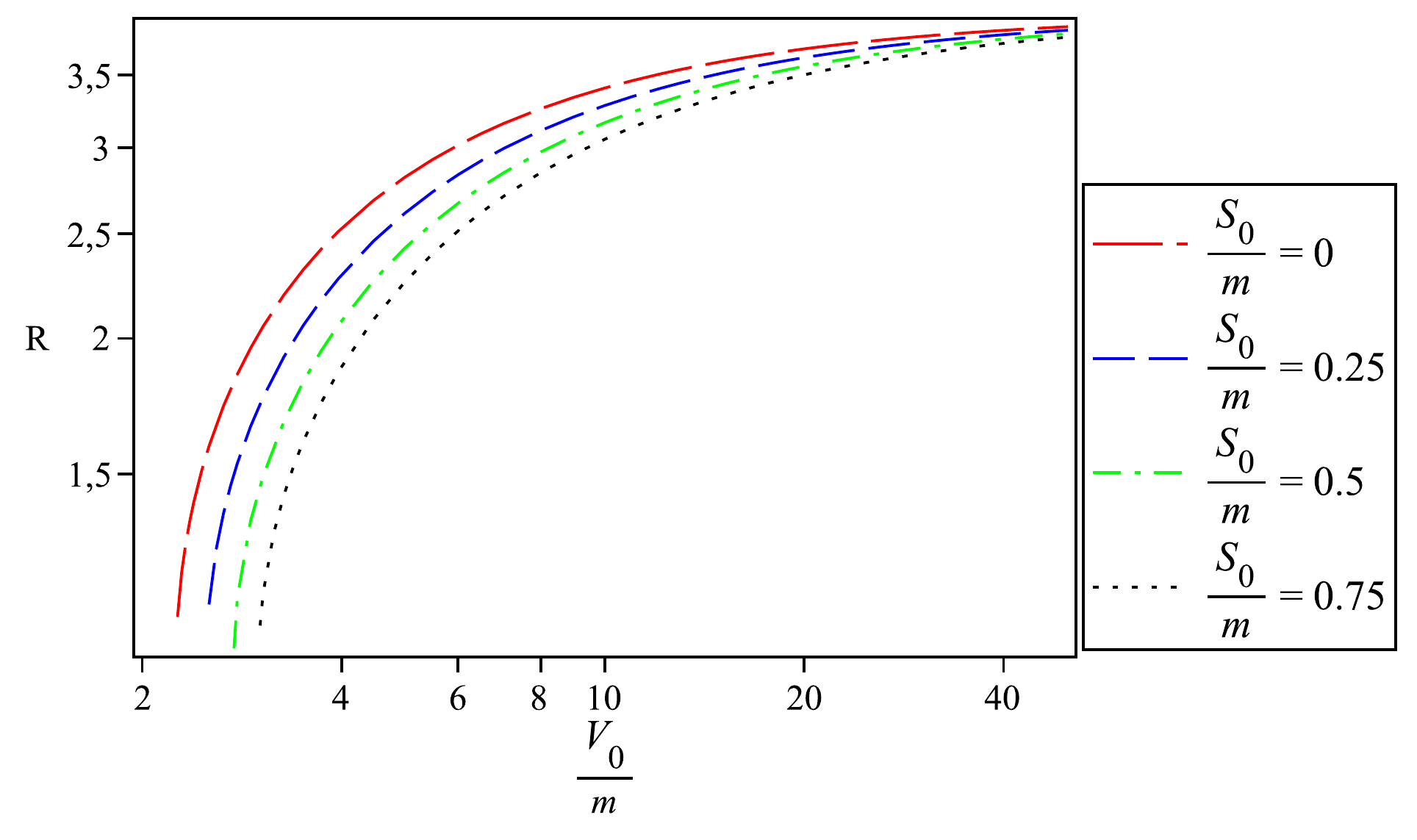}
\caption{Plotting $R$ as function of the variable $\frac{V_{0}}{m}$ for $%
E=1.25m$.}
\label{fig:g4}
\end{figure}
\begin{figure}[H]
\centering
\includegraphics[width=110mm]{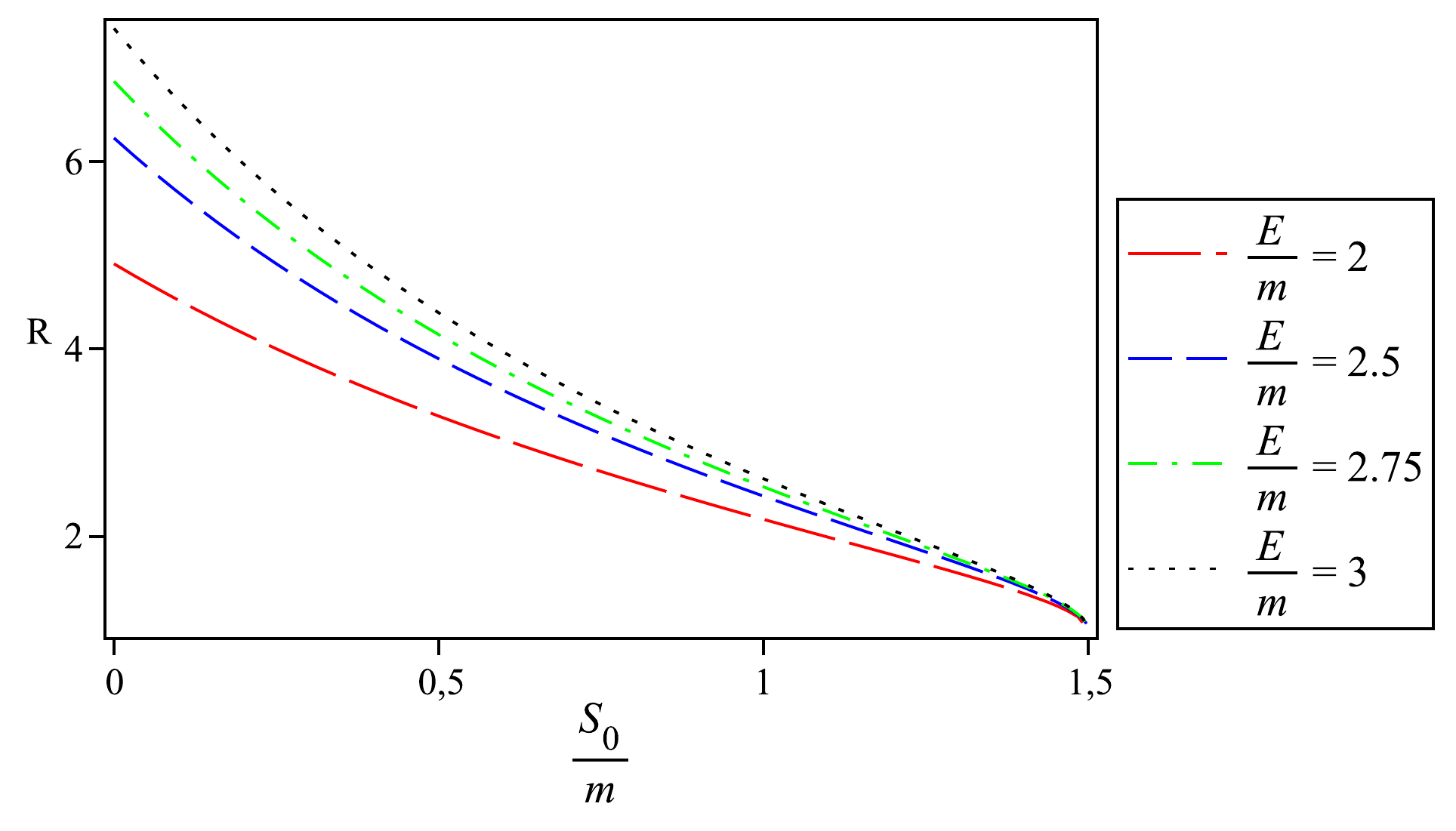}
\caption{Plotting $R$ as function of the variable $\frac{S_{0}}{m}$. $V_{0}$
and $E$ are chosen so that $\frac{V_{0}}{m}-\frac{E}{m}=2.5$.}
\label{fig:g5}
\end{figure}

Plotting $R$ as a function of the variables $\frac{E}{m}$, $\frac{V_{0}}{m}$
and $\frac{S_{0}}{m}$, we show that the considered scalar barrier minimizes
the reflection coefficient in the Klein range. The presence of a scalar
barrier shortens the Klein range. When $S_{0}$ exceeds $V_{0}-2m$, the
Klein's paradox disappears.

In the next section, we consider the Klein's paradox and related particle
creation by a vector and scalar barriers in the framework of quantum field
theory.

\section{Quantum fields interpretation}

Obviously, the treatment of the present problem in the context of
relativistic quantum mechanics has led to paradoxical results. This is
because the sound approach to study quantum processes in the presence of a
potential barrier is rather the quantum field theory in external fields.
Following this approach, we have two different definitions of particles; The
first one is based on the in-basis and the other on the out-basis. These two
definitions are in general different and the difference is a consequence of
the vacuum instability in the presence of the potential barriers. However,
in literature we find two different definitions of the "in" and "out"
states; The first one is suggested by Hansen and Ravndall \cite{Hansen} and
accepted by Greiner and co-workers \cite{Grein3} while the second is
proposed by Nikishov \cite{Nikish2} and has been much used \cite%
{Git2,hao,Gitman,Gitm,Git1,Git3,Git4,Git5}. Recently, a well-built quantum
field theory in the presence of critical potential steps is elaborated in
\cite{Gitman}. In that paper the authors showed that the correct definition
of "in" and "out" states is that of Nikishov. In this section, we follow the
method of \cite{Gitman}.

\subsection{The choice of "in" and "out" states}

Let us first write the solution to the Dirac eqaution in the form%
\begin{equation}
\psi \left( t,z\right) =e^{-iEt}\varphi \left( z\right) ,
\end{equation}%
where $\varphi \left( z\right) $ is a solution of the stationary equation (%
\ref{2}). Besides the solution obtained in the previous section we can
construct other solutions that can be interpreted as the $"in"$ and $"out"$
states of the particle and the antiparticle. We consider only the important
case of the Klein zone where $m<E<V_{0}-m-S_{0}$. In this case, the "in" and
"out" particles are situated on the left of the step and the "in" and "out"
antiparticles are situated on the right of the step. Then the "in" particles
that are moving to the step from the left are subjected to total reflection
and the "in" antiparticles that are moving to the step from the right are
subjected to total reflection. Therefore, acording to \cite{Nikish2} and
\cite{Gitman} the "out" states can be written as follows%
\begin{equation}
\varphi _{out}^{-}\left( z\right) =N_{out}\left\{
\begin{array}{lcc}
u\left( p\right) ~\exp (ipz), &  & z<0 \\
\zeta _{1}v\left( q\right) \exp (iqz)+\zeta _{2}v\left( -q\right) \exp
(-iqz), &  & z>0%
\end{array}%
\right.  \label{d}
\end{equation}%
and%
\begin{equation}
\varphi _{out}^{+}\left( z\right) =N_{out}^{\ast }\left\{
\begin{array}{lll}
v\left( q\right) \exp (iqz), &  & z>0 \\
\zeta _{1}^{+}u\left( p\right) \exp (ipz)+\zeta _{2}^{+}u\left( -p\right)
\exp (-ipz), &  & z<0%
\end{array}%
\right.  \label{f}
\end{equation}%
where%
\begin{eqnarray*}
\zeta _{1} &=&-\zeta _{1}^{+}=-\frac{\left( 1+\gamma _{s}\right) }{2\sqrt{%
\left\vert \gamma _{s}\right\vert }} \\
\zeta _{2} &=&\zeta _{2}^{+}=\frac{1-\gamma _{s}}{2\sqrt{\left\vert \gamma
_{s}\right\vert }}
\end{eqnarray*}%
and the the spinors $u\left( \pm p\right) $ and $v\left( \pm q\right) $ are
given by
\begin{equation}
u\left( \pm p\right) =\sqrt{\frac{E+m}{2p}}%
\begin{pmatrix}
1 \\
\frac{\pm p}{E+m}%
\end{pmatrix}%
\end{equation}%
and%
\begin{equation}
v\left( \pm q\right) =\sqrt{\frac{\left( E-V_{0}\right) +\left(
m+S_{0}\right) }{2\left\vert q\right\vert }}%
\begin{pmatrix}
1 \\
\frac{\pm q}{\left( E-V_{0}\right) +\left( m+S_{0}\right) }%
\end{pmatrix}%
.
\end{equation}%
For the "in" states, we have

\begin{equation}
\varphi _{in}^{-}\left( z\right) =N_{in}\left\{
\begin{array}{lcc}
u\left( -p\right) \exp (-ipz), &  & z<0 \\
\zeta _{2}v\left( q\right) \exp (iqz)+\zeta _{1}v\left( -q\right) \exp
(-iqz), &  & z>0%
\end{array}%
\right.  \label{e}
\end{equation}%
and

\begin{equation}
\varphi _{in}^{+}\left( z\right) =N_{in}^{\ast }\left\{
\begin{array}{lcc}
v\left( -q\right) \exp (-iqz), &  & z>0 \\
\zeta _{2}^{+}u\left( p\right) \exp (ipz)+\zeta _{1}^{+}u\left( -p\right)
\exp (-ipz), &  & z<0%
\end{array}%
\right. .  \label{g}
\end{equation}%
The constants $N_{in}$, and $N_{out}$ are determined according to standard
orthogonality conditions, see Eq. (3.36) in \cite{Gitman},

\begin{eqnarray}
\left( \psi _{in,k}^{\epsilon },\psi _{in,k^{\prime }}^{\epsilon ^{\prime
}}\right) _{z} &=&\epsilon ~\eta _{L}\delta _{\epsilon ,\epsilon ^{\prime
}}\delta _{k,k^{\prime }}  \label{OC1} \\
\left( \psi _{out,k}^{\epsilon },\psi _{out,k^{\prime }}^{\epsilon ^{\prime
}}\right) _{z} &=&\epsilon ~\eta _{R}\delta _{\epsilon ,\epsilon ^{\prime
}}\delta _{k,k^{\prime }}  \label{OC2}
\end{eqnarray}%
where $k$ denotes the conserved quantum numbers (the energy in our case) and
the scalar product of the two states is defined by, see Eq. (3.33) in \cite%
{Gitman},%
\begin{equation}
\left( \psi ^{+},\psi ^{-}\right) _{z}=\int \overline{\psi ^{+}}\boldsymbol{%
\beta \alpha }\psi ^{-}dt.
\end{equation}%
In (\ref{OC1}) and (\ref{OC2}) $\eta _{L}$ and $\eta _{R}$ are defined by $%
\eta _{L,R}=\func{sign}\left( E-V_{L,R}\right) $ with
\begin{equation*}
V_{L}=\lim_{z\rightarrow -\infty }V\left( z\right) =0\text{ \ and \ \ \ }%
V_{R}=\lim_{z\rightarrow +\infty }V\left( z\right) =V_{0}
\end{equation*}%
Tanking into account that

\begin{equation}
\left\vert \zeta _{2}\right\vert ^{2}-\left\vert \zeta _{1}\right\vert ^{2}=1
\end{equation}%
and the spinors $u\left( p\right) $ and $v\left( q\right) $ verify the
following relations
\begin{eqnarray}
\overline{u\left( \epsilon p\right) }~\boldsymbol{\beta \alpha ~}u\left(
\epsilon ^{\prime }p\right) &=&\epsilon ~\delta _{\epsilon ,\epsilon
^{\prime }} \\
\overline{v\left( \epsilon q\right) }~\boldsymbol{\beta \alpha ~}v\left(
\epsilon q\right) &=&\epsilon ~\delta _{\epsilon ,\epsilon ^{\prime }}
\end{eqnarray}%
we find that the orthonormalization conditions (\ref{OC1}) and (\ref{OC2})
are fulfilled for $\left\vert N_{in}\right\vert =\left\vert
N_{out}\right\vert =1$.

Let us note that this definition of the "in" and "out" states in the Klein
zone which coincides with the one proposed by Nikishov is unrelated to
considerations of the previous section.

\subsection{Particle creation in the Klein's range}

Now, according to the standard $S$-matrix formalism of quantum field theory,
the matter field operator admits the following two decompositions
\begin{align}
\mathbf{\psi }& =\underset{k}{\sum }\left( a_{in,k}~\psi
_{in,k}^{+}+b_{in,k}^{+}~\psi _{in,k}^{-}\right)  \label{9} \\
& =\underset{k}{\sum }\left( a_{out,k}~\psi _{out,k}^{+}+b_{out,k}^{+}~\psi
_{out,k}^{-}\right)  \label{10}
\end{align}%
where the operator $a_{in}$ ($a_{out}$) annihilates a particle in the $"in"$
($"out"$) state and the operator $b_{in}^{+}$ ($b_{out}^{+}$) creates an
antiparticle in the $"in"$ ($"out"$) state. These operators verify the
following anticommutation relations

\begin{align}
\left[ a_{in,k},a_{in,k^{\prime }}^{+}\right] _{+}& =\left[
a_{out,k},a_{out,k^{\prime }}^{+}\right] _{+}=\delta _{k,k^{\prime }} \\
\left[ b_{in,k},b_{in,k^{\prime }}^{+}\right] _{+}& =\left[
b_{out,k},b_{out,k^{\prime }}^{+}\right] _{+}=\delta _{k,k^{\prime }}
\end{align}%
and all mixed anticommutators vanish%
\begin{equation}
\left[ a,b\right] _{+}=\left[ a,b^{+}\right] _{+}=0.
\end{equation}%
With these two definitions of particles, the observed created particles are
out-particles in the in-vacuum.

Since the set $\left\{ \psi _{out}^{+}\left( z\right) ,\psi _{out}^{-}\left(
z\right) \right\} $ forms a basis for the solution space of the equation (%
\ref{2}), we can write the elements of the second set $\left\{ \psi
_{in}^{+}\left( z\right) ,\psi _{in}^{-}\left( z\right) \right\} $ as linear
combinations of the functions $\psi _{out}^{+}\left( z\right) $ and $\psi
_{out}^{-}\left( z\right) $. Indeed, taking into account that the spinors $%
u\left( \pm p\right) $ and $v\left( \pm q\right) $ verify the following
relations
\begin{eqnarray*}
v\left( q\right) &=&\zeta _{1}^{+}u\left( p\right) +\zeta _{2}^{+}u\left(
-p\right) \\
v\left( -q\right) &=&\zeta _{2}^{+}u\left( p\right) +\zeta _{1}^{+}u\left(
-p\right) \\
u\left( p\right) &=&\zeta _{2}^{+}v\left( -q\right) -\zeta _{1}^{+}v\left(
q\right) \\
u\left( -p\right) &=&\zeta _{2}v\left( q\right) +\zeta _{1}v\left( -q\right)
\end{eqnarray*}%
we easily establish the developments%
\begin{eqnarray}
\varphi _{in}^{+}\left( z\right) &=&\alpha ^{\ast }\varphi _{out}^{+}\left(
z\right) +\beta ^{\ast }\varphi _{out}^{-}\left( z\right) \\
\varphi _{in}^{-}\left( z\right) &=&\alpha \varphi _{out}^{-}\left( z\right)
+\beta \varphi _{out}^{+}\left( z\right) ,
\end{eqnarray}%
where the Bogoliubov coefficients $\alpha $ and $\beta $ are given by%
\begin{eqnarray}
\alpha ^{\ast } &=&-\frac{\zeta _{1}}{\zeta _{2}}=-\frac{1+\gamma _{s}}{%
1-\gamma _{s}} \\
\beta ^{\ast } &=&\frac{1}{\zeta _{2}}=\frac{2\sqrt{\left\vert \gamma
_{s}\right\vert }}{1-\gamma _{s}}
\end{eqnarray}%
with the condition
\begin{equation}
\left\vert \alpha \right\vert ^{2}+\left\vert \beta \right\vert ^{2}=1.
\label{c}
\end{equation}%
This is the Bogoliubov transformation connecting the "in" with the "out"
states which can be written in the following equivalent form
\begin{eqnarray}
\varphi _{out}^{+}\left( z\right) &=&\alpha \varphi _{in}^{+}\left( z\right)
+\beta \varphi _{in}^{-}\left( z\right) \\
\psi _{out}^{-}\left( z\right) &=&\alpha ^{\ast }\psi _{in}^{-}\left(
z\right) +\beta ^{\ast }\varphi _{in}^{+}\left( z\right) .
\end{eqnarray}%
This relation between "in" and "out" states can be converted to be a
relation between "in" and "out" operators by the use of (\ref{9}) and (\ref%
{10}). We get

\begin{eqnarray}
a_{in} &=&\alpha a_{out}+\beta ^{\ast }b_{out}^{+}  \label{11} \\
b_{in}^{+} &=&\beta a_{out}+\alpha ^{\ast }b_{out}^{+}.
\end{eqnarray}%
We can also write the "out" operators in terms of the "in" operators%
\begin{eqnarray}
a_{out} &=&\alpha a_{in}+\beta ^{\ast }b_{in}^{+}  \label{13} \\
b_{out}^{+} &=&\beta a_{in}+\alpha ^{\ast }b_{in}^{+}.
\end{eqnarray}%
Here, we notice that that the obtained Bogoliubov coefficients completely
determine the quantum processes in the presence of a scalar barrier in
addition to the usual vector barrier. This procedure allows us to calculate
all physical quantities in a simple way. For instance, to deal with the
particle creation process, we consider the amplitude

\begin{equation}
\mathcal{A}=\left\langle 0_{out}\left\vert a_{out}b_{out}\right\vert
0_{in}\right\rangle
\end{equation}%
and by taking into account that\qquad

\begin{equation}
b_{out}=\frac{1}{\alpha ^{\ast }}b_{in}+\frac{\beta ^{\ast }}{\alpha ^{\ast }%
}a_{out}^{+}
\end{equation}%
we find%
\begin{equation}
\mathcal{A}=-\frac{\beta ^{\ast }}{\alpha ^{\ast }}e^{iw},
\end{equation}%
where $e^{iw}$ is the vacuum to vacuum amplitude
\begin{equation}
e^{iw}=\left\langle 0_{out}\right\vert \left. 0_{in}\right\rangle .
\end{equation}%
The absolute probability to create particles in the vicinity of the barrier
is then
\begin{equation}
P_{Cr}=\left\vert \mathcal{A}\right\vert ^{2}=\left\vert \frac{\beta ^{\ast }%
}{\alpha ^{\ast }}\right\vert ^{2}C_{0}=P_{rel}~C_{0}
\end{equation}%
where $C_{0}$ is the vacuum to vacuum probability%
\begin{equation}
C_{0}=\left\vert \left\langle 0_{out}\right\vert \left. 0_{in}\right\rangle
\right\vert ^{2}
\end{equation}%
and $P_{rel}$ is the relative probability to create a pair
\begin{equation}
P_{rel}=\left\vert \frac{\beta ^{\ast }}{\alpha ^{\ast }}\right\vert ^{2}=%
\frac{4\left\vert \gamma _{s}\right\vert }{\left( 1+\gamma _{s}\right) ^{2}}
\end{equation}%
Taking into account the Pauli exclusion principle, we have
\begin{equation}
C_{0}+\left\vert \frac{\beta ^{\ast }}{\alpha ^{\ast }}\right\vert
^{2}C_{0}=1,
\end{equation}%
which gives, directly, the vacuum persistence (vacuum to vacuum probability)
\begin{equation}
C_{0}=\frac{1}{1+\left\vert \frac{\beta ^{\ast }}{\alpha ^{\ast }}%
\right\vert ^{2}}=\left\vert \alpha \right\vert ^{2}=\left( \frac{1+\gamma
_{s}}{1-\gamma _{s}}\right) ^{2}.
\end{equation}%
We, therefore, have%
\begin{equation}
P_{Cr}=\left\vert \frac{\beta ^{\ast }}{\alpha ^{\ast }}\right\vert
^{2}~\left\vert \alpha \right\vert ^{2}=\left\vert \beta \right\vert ^{2}=%
\frac{4\left\vert \gamma _{s}\right\vert }{\left( 1-\gamma _{s}\right) ^{2}}=%
\frac{T}{R}.
\end{equation}%
Another important result of the Pauli principle is that only one pair could
be created in well-defined state. The mean number of created particles in
the state $E$ is then%
\begin{equation}
\overline{n}=P_{rel}~C_{0}=\left\vert \beta \right\vert ^{2}=\frac{%
4\left\vert \gamma _{s}\right\vert }{\left( 1-\gamma _{s}\right) ^{2}},
\end{equation}%
which is the same as the mean number of "out" particles in the "in" vacuum $%
\overline{n}=\left\langle 0_{in}\left\vert a_{out}^{+}a_{out}\right\vert
0_{in}\right\rangle =\left\vert \beta \right\vert ^{2}$.

\begin{figure}[H]
\centering
\includegraphics[width=110mm]{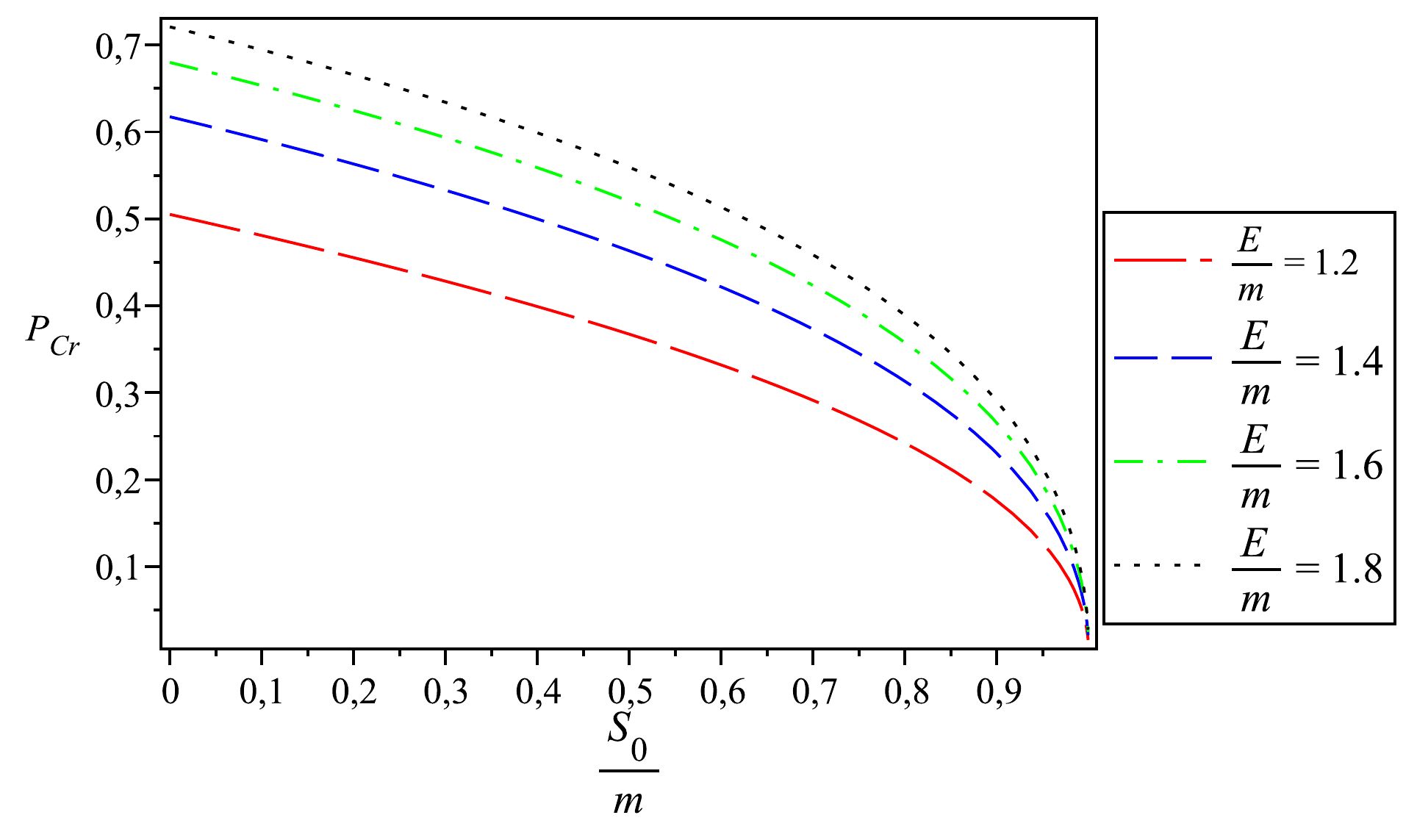}
\caption{Plotting the probability $P_{Cr}$ as a function of the variable $%
\frac{S_{0}}{m}$ for several values of $\frac{E}{m}$.}
\label{fig:pcr1}
\end{figure}

In Figure (\ref{fig:pcr1}) we plot the absolute probability $P_{Cr}$ as a
function of the variable $\frac{S_{0}}{m}$ for several values of $\frac{E}{m}
$ and $\frac{V_{0}}{m}$ with $V_{0}=E+2.5m$. The result is as expected, the
scalar barrier, as considered in this paper, minimizes the absolute particle
creation probability.

\begin{figure}[H]
\centering
\includegraphics[width=110mm]{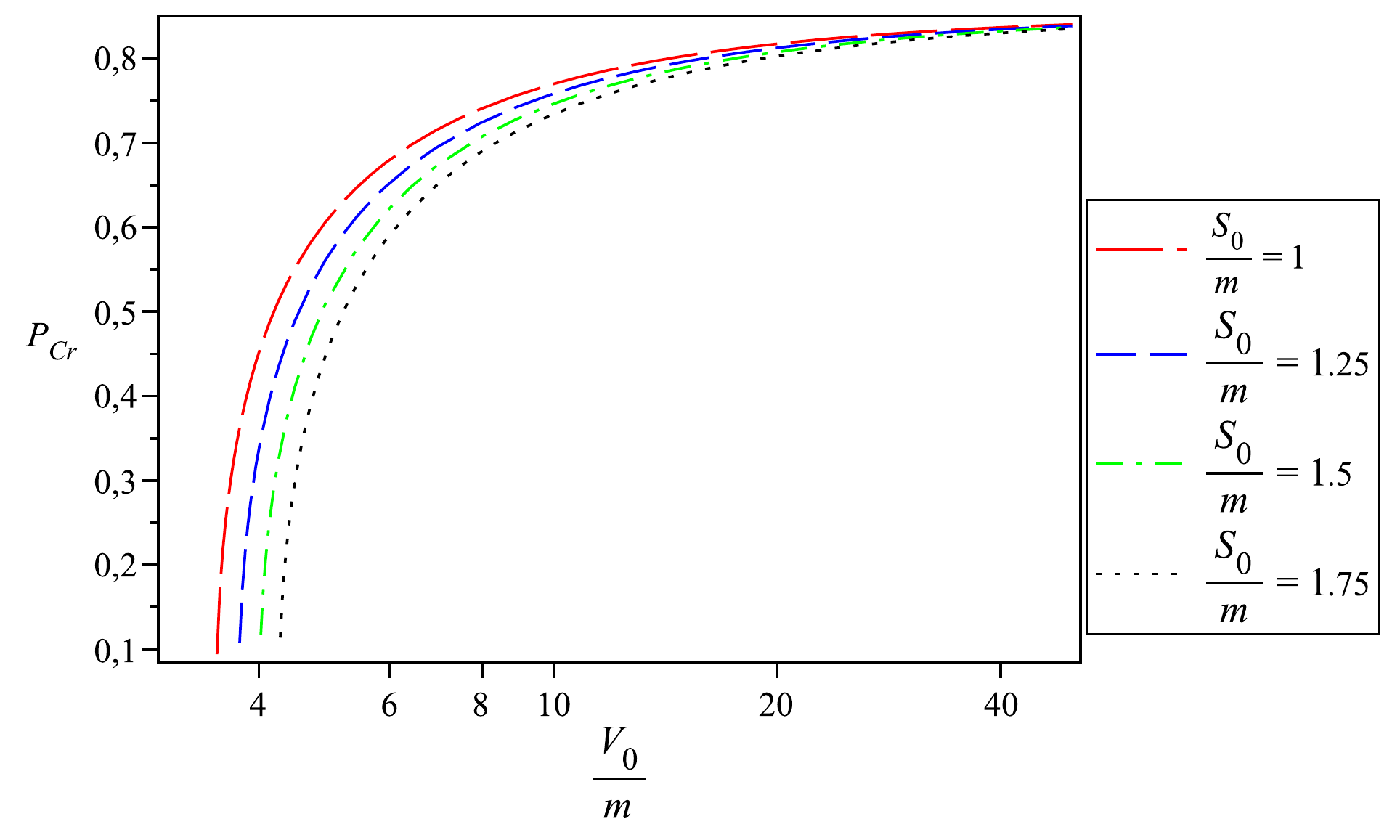}
\caption{Plotting the probability $P_{Cr}$ as a function of the variable $%
\frac{V_{0}}{m}$ for several values of $\frac{S_{0}}{m}$.}
\label{fig:pcr2}
\end{figure}

In Figure (\ref{fig:pcr2}) we plot the absolute probability $P_{Cr}$ as a
function of the variable $\frac{V_{0}}{m}$ for several values of $\frac{S_{0}%
}{m}$ and $\frac{V_{0}}{m}$ with $V_{0}=E+2.5m$. In addition to reducing the
particle creation, plots in Figure (\ref{fig:pcr2}) show also that as $V_{0}$
increase as the effect of the scalar barrier becomes inessential.

\section{Conclusion}

We have studied, in this paper, the Klein's paradox in the presence of
scalar and vector potential barriers. At the first stage, we have found the
exact solutions of the corresponding Dirac equation. Then, we have extracted
form these solutions the transmission and reflection coefficients. It is
shown that the presence of a scalar barrier in addition to the vector one
minimizes the reflection coefficient in the Klein range. The presence of a
scalar barrier shortens the Klein range. When $S_{0}$ exceeds $V_{0}-2m$,
the Klein's paradox disappears.

In order to get the good interpretation of the pair creation we have
considered a field theoretical treatment following the rigorous theory
elaborated in \cite{Gitman}. At the first stage, we have defined the "in"
and "out" states. Then, we were able to extract the Bogoliubov coefficients
and to calculate the pair production probability and the mean density of
created particles. We have shown that the scalar potential decreases the
relative probability to create a pair of particles. This result can be
explained by the fact that the gap between positive and negative energies in
the region $(z>0)$ in the presence of the vector potential is $2m$ while
this gap becomes \ $2m+2S_{0}$ in the presence of the scalar potential. This
means that, in the presence of the scalar potential, the forbidden band to
the right of the barrier is larger than the usual one and therefore, the
scalar potential cut down the Klein range and minimizes the creation of
particles. The particle creation decreases as the scalar potential increases
and ceases definitely when the scalar potential reaches the value $V_{0}-2m$%
. In other words, in the presence of scalar and vector barriers, only
particles of mass $m$, with $2m<V_{0}-S_{0}$, can be created.

Theoretically, this result could have a significant impact in nuclear
physics; In the case of spin and pseudospin symmetries where the magnitudes
of the scalar and vector potentials are equal \cite%
{Ginocchio1,Ginocchio2,Ginocchio3}, the Klein range would be suppressed by
the scalar potential and the creation of particles by the tunneling
mechanism would then be impossible.

\end{document}